# Negative compressibility in MoS$_2$ capacitance

Ruiyan Gao, Zhehan Ying, Liheng An, Zefei Wu, Xiangbing Cai, Shi Wang, Ziqing Ye, Xuemeng Feng, Meizheng Huang, and Ning Wang*

*Department of Physics and the William Mong Institute of Nano Science and Technology, The Hong Kong University of Science and Technology, Clear Water Bay, Kowloon, Hong Kong, China*

**Abstract:** Large capacitance enhancement is useful for increasing the gate capacitance of field-effect transistors (FETs) to produce low-energy-consuming devices with improved gate controllability. We report strong capacitance enhancement effects in a newly emerged two-dimensional channel material, molybdenum disulfide (MoS$_2$). The enhancement effects are due to strong electron-electron interaction at the low carrier density regime in MoS$_2$. We achieve about 50% capacitance enhancement in monolayer devices and 10% capacitance enhancement in bilayer devices. However, the enhancement effect is not obvious in multilayer (layer number >3) devices. Using the Hartree-Fock approximation, we illustrate the same trend in our inverse compressibility data.

Negative compressibility occurs in a strong electron-electron interaction electronic system. When increasing the carrier density of the system, the chemical potential of the system will be lowered. This effect can be reflected in the enhancement of capacitance [1-4], providing a large capacitance surge despite the limited geometric cross-sectional area of the gate in two-dimensional (2-D) electron gas. The modulation of this kind of enhancement of capacitance could be quite useful in increasing the gate capacitance of FETs. Similar effects have been previously discovered in ultraclean heterostructures like $LaAlO_3/SrTiO_3$ interfaces[5], GaAs quantum wells[3,4,6,7], atomic thin black phosphorus (BP)[8], and graphene[9]. Here, we report that atomically thin $MoS_2$, a newly emerged 2-D semiconductor for FET channel material, has high capacitance enhancement effects at low carrier density within the low frequency range.

Atomically thin $MoS_2$ capacitors were fabricated based on hexagonal boron nitride (h-BN) encapsulated device structure. The employment of a dry-transfer method offers a valid precondition for producing an ultraclean interface and polymer-free environment on the BN-$MoS_2$-BN capacitance heterostructures. This method has also be used for investigating the quantum Hall effect[10] and other transport phenomena of charge carriers in 2-D materials[11-14]. Atomically thin $MoS_2$ flakes were exfoliated from bulk crystal with tapes and transferred onto p-Si substrates with a 300nm thick layer of thermal silicon dioxide ($SiO_2$). The back electrodes were adopted as surface contact terminal electrodes. The original flakes were exfoliated from bulk crystal which has fewer defects than chemical vapor deposition (CVD) products. Annealing at 300 °C (1 hour) was carried out to squeeze out small air bubbles that were induced during the fabrication procedure. Standard electron lithography and reactive-ion etching are employed to make FET devices. Then, a Ti/Au(5/65nm) top gate was deposited while carefully avoiding the bubbles which cannot be eliminated by the annealing process. One of the devices is shown in the optical microscope image in Fig. 1(c) and the diagram (Fig. 1(b)) yields the basic structure of the samples. All of our measurements of the negative capacitance were carried out at room temperature using an HP Precision E4980A LCR meter with a sensitivity of ~0.1fF.

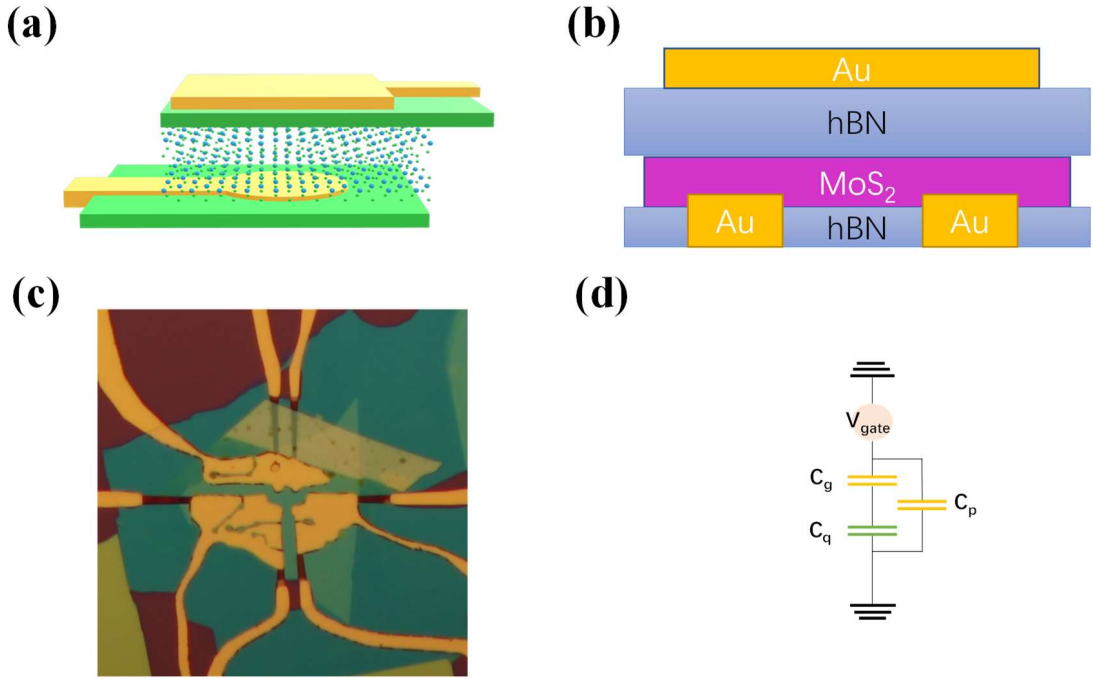

**Fig. 1**. MoS$_2$ back electrode capacitor structure (a)(b) schematic imagine of BN-MoS$_2$-BN structure with back electrode. The thickness of MoS$_2$ is less than 1nm as determined using atomic force microscope. (c) the optical imagine of one of our devices. (d) the equivalent circuit of MoS$_2$ capacitance devices. the total capacitance can be regarded as the serial connect of geometric capacitance and quantum capacitance. Meanwhile in circuit there is inevitable parasite capacitance.

Unlike the metal oxide dielectric layer in MOSFET[15], the dielectric layer we use in our samples is h-BN, which has an atomically uniform flat surface and offers a stable dielectric property that depends on frequency. Meanwhile, we deposited Ti/Au(5/20nm) on the etched h-BN channel to form the back electrode and the large area back electrode ensures good contact with monolayer MoS$_2$. The structure of BN- MoS$_2$-BN devices can be simplified to a serial connecting capacitor circuit of geometric capacitance and quantum capacitance.

This device can be separated into two parts(Fig. 1(d)), one is geometric capacitance $C_g = \frac{\varepsilon_{BN}}{d}$(per area) and other is quantum capacitance[16,17] $C_q = e^2 \frac{dn}{d\mu}$ where d is the distance between the top gate and channel material, e is the elementary charge, and $\frac{dn}{d\mu}$ is the compressibility of the carrier density[18,19], defined as the change of carrier density $n$ over unit variation of chemical potential $\mu$. The existence of parasitic capacitance could affect our measurement results, however, the parasitic capacitance in our sample is in the same order of 1$fF$ in the circuit which was measured using a gate without channel material. The compressibility describes the ability to screen the external charges. For metals, this compressibility usually approaches infinity. The total capacitance can be simplified to $\frac{1}{C_t} = \frac{1}{C_g} + \frac{1}{C_q}$. The Fermi energy can be modulated by changing the top gate voltage. Continuously changing the Fermi energy results in different compressibility, which is proportional to the density of states at the Fermi level, thus changing the quantum capacitance. When the total capacitance is larger than the geometric one, negative compressibility occurs.

Raman spectra were obtained to determine the layer number of MoS$_2$ in the device. The crystalline 2H- MoS$_2$ belongs to the $D_{6h}$ crystal class and by determining the Raman band positions of the $E_{2g}^1$ and $A_{1g}$ the layers of sample can be determined. Raman spectra shown in Fig. 2(b)) indicate an in-plane $E_{2g}^1$ vibrational mode and out-of-plane $A_{1g}$ vibrational mode. The in-plane mode corresponds to the sulfur atoms vibrating in one direction and the molybdenum atom in the other. The separation between these two vibrational modes indicates the number of layers of the sample. As the thickness increases in MoS$_2$ flakes the in-plane mode upshifts and the out-of-plane downshifts. For sample 1, the separation is 23 cm$^{-1}$, which shows that this is a monolayer sample[20].

The devices were measured in vacuum, and the excitation frequency was from 100kHz to 1MHz. Figure 3 shows the capacitance measured at room temperature. After the depletion region, the large capacitance enhancement occurs in the low carrier density region. For the capacitance measurement in MoS$_2$ samples, the frequency should be carefully chosen. The capacitor may not be fully charged if the frequency is $f > \frac{1}{RC_t}$ where R is the lateral resistance which needs to be considered[12]. Large enhancement of capacitance can only be observed in the relatively low frequency measurements since high frequency measurement may lead to the vanishing of some states and the result of experiments indicates a relatively low capacitance.

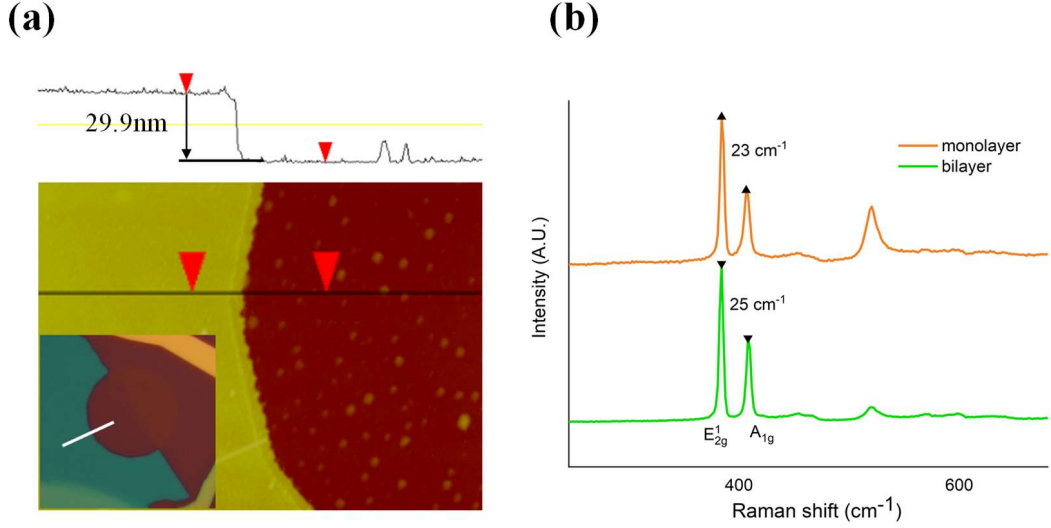

**Fig. 2**. (a) the determining of thickness of hBN in our monolayer samples which is 29.9 nm. (b) the layer of MoS$_2$ channel material was also confirmed via Raman spectroscopy.

All of our samples show sharp capacitance diminishment near the depletion region. The relationship between carrier density and gate voltage can be attained from $n = \frac{C_g}{e}(V_g - V_{th})$, where $V_{th}$ is the threshold voltage. For a monolayer device, the sample becomes conductive when the top gate reaches Vg= -0.15 V. The capacitance enhancement happens when the frequency is lower than 500kHz as indicated by the peak near the depletion region (Fig. 3(a)). When the voltage of the gate is larger than 2V (the corresponding carrier density $n = 1.36 \times 10^{12} cm^{-2}$), the channel material is fully charged and the capacitance can be regarded as geometric capacitance $C_g = A\frac{\varepsilon}{d}$, where $A$ is the overlap area of the metal gate and the conductive channel, $d$ is the thickness of the dielectric layer. Usually, the capacitance at a high carrier density $C_{hd}$ is nearly $C_g$ because the difference between these two values is quite small which can be ignored. Based on this assumption, the capacitances in the high-density region are treated as the geometric capacitances of the samples in this paper. The curve shows an upturn trend at high carrier density, but at high frequency this upturn trend disappears. In this device, the enhancement of capacitance is about 51% larger than the capacitance in high carrier density at 100kHz, and this enhancement has a quick decay with increasing frequency. For the region in which the frequency is higher than 800kHz, this enhancement totally disappears.

We did not detect any capacitance enhancement near the depletion in multilayer (>3) MoS$_2$ capacitance samples as shown in Fig. 3(b). We noticed the turn-on voltages of these thicker samples ($V_{gate}^{turnon} \approx -2.5V$ for this sample and for another sample $V_{gate}^{turnon} \approx -4V$) are far less than those of monolayer samples. The reason that the enhancement is weaker for a bilayer is that the electron motions are not strictly in 2-D movement with finite thickness. A simple explanation is that electrons' interaction is much weaker in 3D. Interaction effects are not apparent unless the density is low. However, the observed difference between low and high frequency signal data in a multilayer suggests that the capacitance difference is induced by charge traps[12]. Actually, the charge traps can be easily induced by impurities and defects in the channel material and could be excited at relatively low frequencies. In our multilayer samples, the traps' induced capacitance difference is less than 10fF which is much small than the capacitance enhancement in the monolayer samples. However, Chen, *et. al.,* reported that the resonant scattering that is caused by impurities could enhance negative compressibility[9].

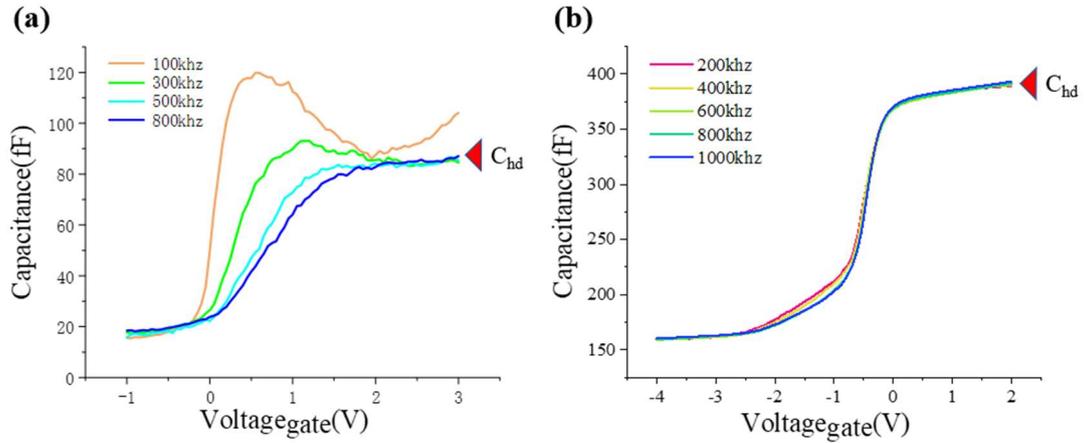

**Fig. 3**. the capacitance measurements which are carried out in room temperature. (a) monolayer MoS$_2$ capacitance devices showed that the total capacitance considerably exceeds the capacitance in high carrier region $C_{hd}$. The enhancement can reach 51%. (b) multilayer sample shows no evidence of negative compressibility but charge traps, which locates in $-2.5V \leq V_g \leq -0.5V$.

Measuring negative capacitance requires an ultraclean heterostructure interface and low density of defects in the samples. Presumably, the sharpness of the upturn trend and the amplitude of the enhancement is limited by the local density of states induced by defects and metal-insulator transition[5]. In general, our monolayer samples' data reflect the high quality in the interface of the heterostructure.

**Hartree-Fock approximation and inverse compressibility data**

The geometric capacitance and quantum capacitance models yield the total capacitance which normally could not be larger than the geometric capacitance.

However, if the compressibility is negative in the system, this kind of large enhancement could happen. Here, we use inverse compressibility $\frac{d\mu}{dn}$ to describe the strong electron-electron interaction in the 2-D system.

In an ideal 2-D electron gas system, the ground state energy consists mainly of two parts. One is the average kinetic energy $E_k = \frac{n\pi\hbar^2}{m^*}$, the other is the electron exchange energy $E_{ex} = -(2/\pi)^{1/2}\frac{e^2}{4\pi\varepsilon}n^{1/2}$, where $m^*$ is effective mass and $\varepsilon$ is the relative dielectric constant of the electron system. The total energy per unit area is $E_t^{HF} = \left(\frac{1}{2}E_k + E_{ex}\right)n$, where $n$ is the carrier density. By definition, the inverse compressibility is $K^{-1} = n^2 \frac{\partial^2 E_t^{HF}}{\partial n^2} = n^2 \frac{\partial \mu}{\partial n}$, hence $\frac{\partial^2 E_t^{HF}}{\partial n^2} = \frac{\partial \mu}{\partial n}$. Thus, for a homogeneous 2-D electron gas, the inverse compressibility is given by[21]

$$\frac{\partial \mu}{\partial n} = \frac{\hbar^2 \pi}{m^*} - \left(\frac{2}{\pi}\right)^{1/2}\frac{e^2}{4\pi\varepsilon}n^{-1/2}$$

The exchange energy could strongly affect the electron compressibility when the carrier density is low enough, and if the carrier density approaches zero, the compressibility could reach negative infinity theoretically. However, when the carrier density increases, the compressibility will approach a constant value.

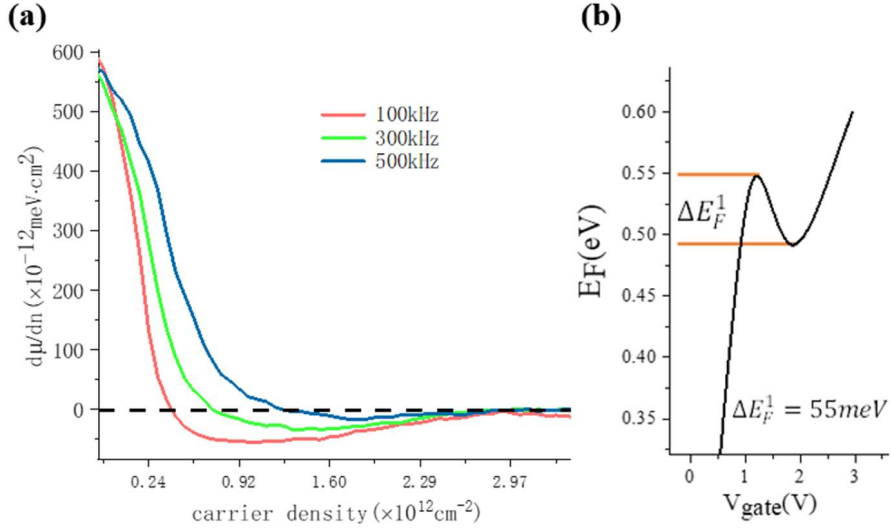

**Fig. 4**. inverse compressibility and Fermi energy (a) the inverse compressibility of sample 1. Negative inverse compressibility occurs in the region of positive carrier density which indicates the carrier is electron. (b) the fermi energy can be calculated by integrating the $V_s \sim V_g$ relation. The $\Delta E_F$ in reversed region is 55meV.

In our samples, we observe that the compressibility curve drops sharply through zero when the channel material becomes conductive. The compressibility is negative through a large region of carrier density. The positive sign of carrier density indicates the nature of electron carrier. As the frequency increases, the negative compressibility is suppressed, and the data show a normal value approaching zero when the electron gas forms. At the lowest point of compressibility, the corresponding carrier density is $0.95 \times 10^{11}$ cm$^{-2}$ and the mean interparticle distance is $r_s = 30$. However, for multilayer devices (more than 3 layers), no negative compressibility is observed. The Fermi energy $E_F$ can be calculated from $E_F = eV_s$, where $V_s$ is the surface potential of MoS$_2$, which can be extracted from $V_s = \int_0^{V_g}(1 - \frac{C_t}{C_g})dV_g$. There is an upturn showing in the Fermi energy curve in monolayer samples, which indicates the signature of negative compressibility. After all, the simple Hartree-Fock approximation shows us the right trend of change of compressibility versus carrier density, which indicates there are strong electron-electron interactions in our 2-D electron system.

**Conclusion**

In conclusion, we successfully fabricated FETs with atomically thin MoS$_2$ as the channel material and h-BN as the dielectric layer. These FETs show extremely large capacitance enhancement near the depletion region at low frequency. We investigated the strong electron-electron interaction mechanism in the FETs using the Hartree-Fock approximation explain the inverse compressibility data we observed. The negative capacitance effect is much weaker in bilayer samples and disappears in multilayer samples because the electron motions are not strictly 2-D with non-zero thickness are not ideal 2D electron gas. Strong electron-electron interaction in low carrier density cannot be maintained when the layer number is larger than 3.

**Conflict of interest statement**
The authors declare that they have no potential conflict of interest to this work.

**Acknowledgments**
This work is supported by the Research Grants Council of Hong Kong (Project No. 16300717 and C7036-17W). We also acknowledge the technical support from the Super-resolution Electron Microscopy facility (C6021-14E) and Raith-HKUST Nanotechnology Laboratory for the electron-beam lithography facility at MCPF.

# Figure Captions

**Fig. 1**. MoS$_2$ back electrode capacitor structure (a)(b) schematic imagine of BN-MoS$_2$-BN structure with back electrode. The thickness of MoS$_2$ is less than 1nm as determined using atomic force microscope. (c) the optical imagine of one of our devices. (d) the equivalent circuit of MoS$_2$ capacitance devices. the total capacitance can be regarded as the serial connect of geometric capacitance and quantum capacitance. Meanwhile in circuit there is inevitable parasite capacitance.

**Fig. 2**. (a) the determining of thickness of hBN in our monolayer samples which is 29.9 nm. (b) the layer of MoS$_2$ channel material was also confirmed via Raman spectroscopy.

**Fig. 3**. the capacitance measurements which are carried out in room temperature. (a) monolayer MoS$_2$ capacitance devices showed that the total capacitance considerably exceeds the capacitance in high carrier region $C_{hd}$. The enhancement can reach 51%. (b) multilayer sample shows no evidence of negative compressibility but charge traps, which locates in $-2.5V \leq V_g \leq -0.5V$.

**Fig. 4**. inverse compressibility and Fermi energy (a) the inverse compressibility of sample 1. Negative inverse compressibility occurs in the region of positive carrier density which indicates the carrier is electron. (b) the fermi energy can be calculated by integrating the V$_s$~V$_g$ relation. The $\Delta E_F$ in reversed region is 55meV.

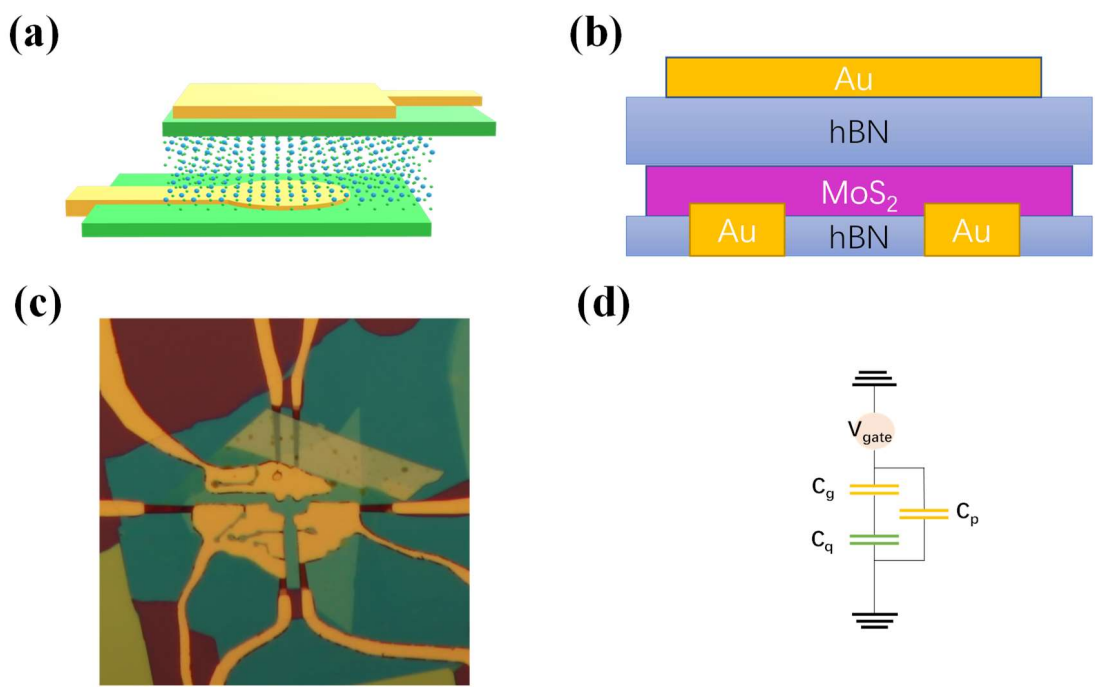

**Fig. 1**

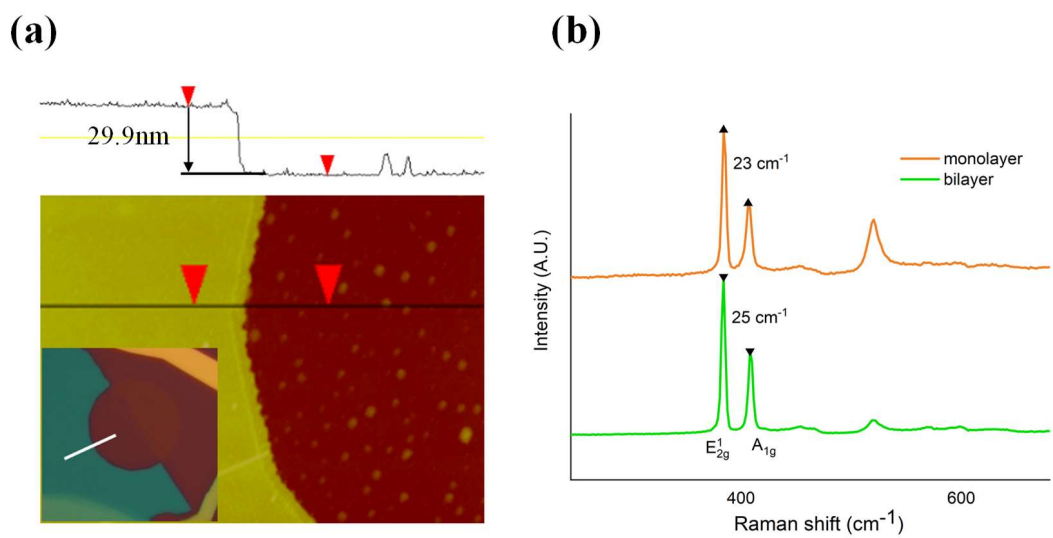

**Fig. 2**

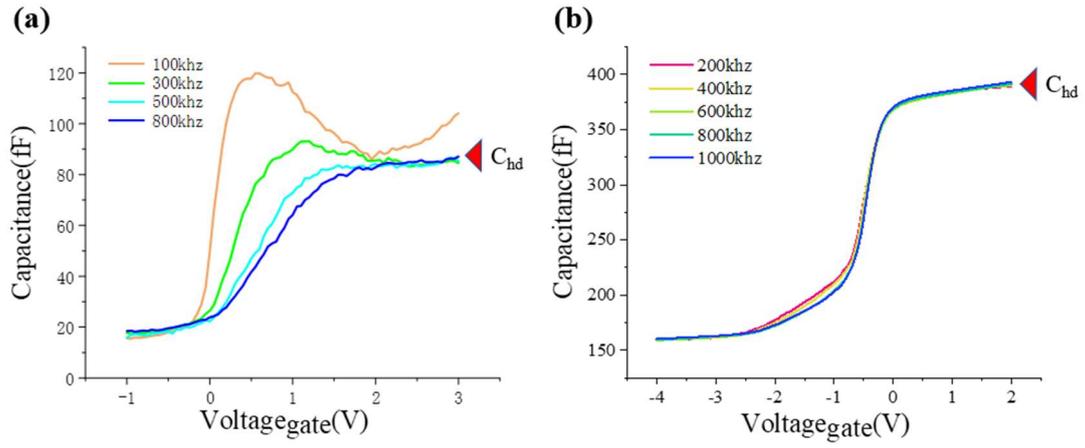

**Fig. 3**

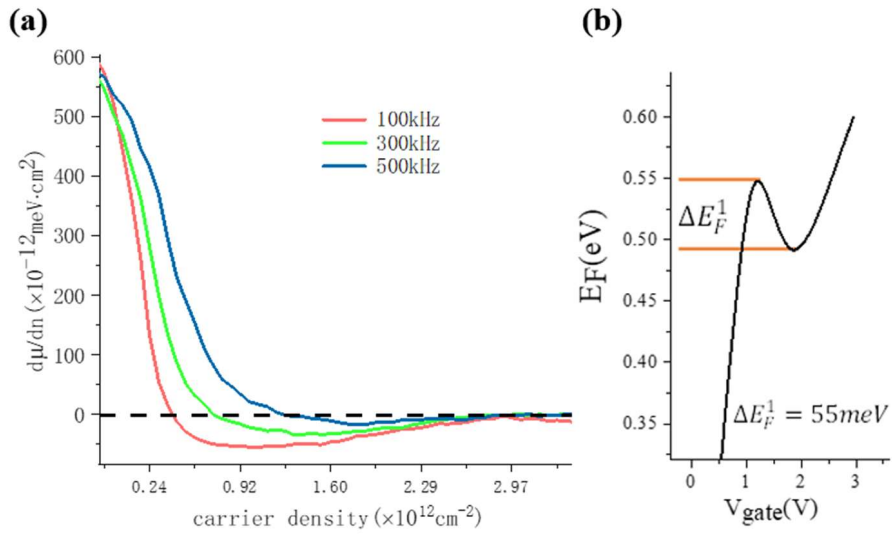

**Fig. 4**